# Study of surface-bulk mass transport and phase transformation in nano TiO$_2$ using hyperfine interaction technique

S.K. Das*, D. Banerjee, S.V. Thakare[1], P.Y. Nabhiraj[2], R. Menon[2], R.K. Bhandari[2], K. Krishnan[3]

Radiochemistry Laboratory, Variable Energy Cyclotron Centre
Bhabha Atomic Research Centre, 1/AF Bidhan nagar, Kolkata 700064, India
[1]Radiopharmaceuticals Division, Bhabha Atomic Research Centre, Mumbai 400085, India
[2]Variable Energy Cyclotron Centre, 1/AF Bidhan nagar, Kolkata 700064, India
[3]Fuel Chemistry Division, Bhabha Atomic Research Centre, Mumbai 400085, India

**Abstract**

Phase transition from anatase to rutile for the 70nm TiO$_2$ crystallite has been investigated by annealing at different temperatures followed by TDPAC measurement of these TiO$_2$ crystallites adsorbed with $^{181}$Hf tracer. The width of the peaks in XRD spectra for TiO$_2$ crystallites annealed at different temperatures indicates the growth of the crystallites with temperature. The samples annealed upto 823K for 4hrs showed no phase transition, except the growth of the crystallites. However, it showed phase transition at the same temperature (823K), when annealed for longer duration. Annealing for shorter duration at 1123K showed phase transition. $^{181}$Hf tracer adsorbed on 70 nm anatase TiO$_2$ is found to be in different polymorphs (anatase & rutile) formed during annealing process. Surface to bulk mass-transfer is found to play a significant role in the phase transition process.

## Introduction

In last two decades, $TiO_2$ has emerged as one of the most important materials which has attracted research of basic nature [1] and found applications in diverse and technologically important areas [2-4]. $TiO_2$, either pure or in doped form, has uses as photocatalyst, energy converter in solar cells, white pigment in paints, sunscreen material, just to name a few. Thin film of $TiO_2$ as an upcoming spintronics would find technological application in electronic industry [5].

$TiO_2$ among several natural polymorphs exists in two main crystalline modifications- anatase and rutile. Anatase is kinetically while rutile is thermodynamically stable. The temperature range at which the phase transition from anatase to rutile takes place is 673K -1473K [6] for bulk $TiO_2$. The transformation from rutile to anatase for the bulk is forbidden. But at nano dimension (<16nm) phase transformation reverses from rutile to anatase[7]. This is due to the higher surface energy of rutile than anatase(1.32 $J.m^{-2}$ for anatase and 1.91 $J.m^{-2}$ for rutile). Therefore, the stability of either polymorph has important bearing on the size of the crystallite. This aspect of the average size dependence on the anatase to rutile transformation has been studied by several authors [8-11].

In different applications, it is either anatase or rutile form which is required. For example, in photocatalysis [12], anatase shows higher activity than rutile. Since the photocatalysis process occurs on the surface, crystallites with low dimension are more useful. Therefore it is important to understand the nature of surface and phase stability with respect to the size of the crystallite. In certain situation, namely, in nuclear waste disposal [13], high temperature sintering of $TiO_2$ to make the waste form is required. This process is enhanced by phase transformation which again has strong dependence on the crystallite size.

It is thus evident that anatase to rutile phase transformation is an important aspect of $TiO_2$ and this is required to be understood to use $TiO_2$ in different applications. Various techniques including XRD, TEM techniques [6, 14] have been used to understand this process.

The probable mechanism for the anatase to rutile transformation was suggested as the nucleation of rutile on the coarser $TiO_2$ ( anatase ) due to the cooperative movement of Ti

& O atoms [6]. The observation of the growth of rutile polymorph during the transformation was explained as the formation of rutile crystal and its rotation around another particle either anatase or rutile to convert the smaller rutile particle into a bigger rutile particle when they attach each other in a favorable crystallographic orientation.

In this paper we have addressed the issues related to the role played by the coarsening of smaller crystallites and also the surface/bulk mass transfer in the temperature mediated transformation process. We have used the nuclear technique TDPAC with atomic scale resolution to study these processes. We first measured the hyperfine parameters for the bulk anatase and rutile forms of $TiO_2$ [15]. The measurement for the rutile is a repetition except the difference in the preparation of the sample mentioned elsewhere [16]. The parameters thus obtained after annealing the sample at different temperatures have been used to address surface/bulk mass transfer and the different aspects of the phase transformation taking place in 70 nm $TiO_2$.

**Experimental**

Sample preparation:

i) $TiO_2$ doped with $^{181}Hf$:

The method of preparing $TiO_2$ and doping with nuclear probe has been described in our previous work [17]. The only difference is that we are doping $TiO_2$ with $^{181}Hf$ in the present work in place of $^{44}Ti$ used in our previous work. The sample was annealed at different temperatures and counted on the coincidence setup mentioned below.

ii) $TiO_2$ adsorbed with $^{181}Hf$:

Commercially available $TiO_2$ in the anatase form was ball milled to bring down the average size of $TiO_2$ to ~70 nm. This size and its distribution were confirmed by AFM and XRD.

A stock solution of $^{181}Hf$ radioactivity was made by dissolving the neutron irradiated $HfCl_4$ in water. Adsorption of $^{181}Hf$ activity on the above mentioned nano crystallites of $TiO_2$ was carried out in pH=7 solution from solid solution interface by mixing weighed (300 mgs) amount of the nano $TiO_2$ in 25 ml of solution containing the $^{181}Hf$ tracer. This mixture was stirred continuously for two hours at room temperature to reach the equilibrium. The $^{181}Hf$ labeled $TiO_2$ was then washed with water several times and dried at 373K to remove any moisture from TiO$_2$. After counting, the same sample was annealed at 823K, 1123K for different duration and was counted on TDPAC spectrometer after annealing.

We observed in our previous experiment [17] that the TiO$_2$ prepared by hydrolysis of Ti-isopropoxide showed the transformation temperature for the anatase to rutile around 1123K. This temperature varies depending on the nature of sample. The transition temperature for the nano sized particles is below that for the bulk material. The second set of samples consisted of the rutile samples prepared by heating the starting anatase TiO$_2$ at 1123K. $^{181}$Hf tracer was then adsorbed on this rutile sample by the same method described above. This sample was then counted for the TDPAC measurement. The third set of samples consisted of two samples prepared by adsorbing $^{181}$Hf activity on the 70 nm anatase modification. These samples were annealed at 823K, 1123K respectively for 8 hours. One of these samples annealed at 823K was once again annealed at 1223K for 8 hours and counted.

**TDPAC measurement**

Fast-slowcoincidence of 133 – 482 Kev cascade of the probe $^{181}$Ta was measured using a coincidence system consisting of three BaF$_2$ detectors. The coincidence data acquisition was performed in list mode by the help of CAMAC electronics mentioned in ref [15]. Individual time spectra at 90$^0$ and 180$^0$ angles were obtained by applying the cascade energy gates through software in the post acquisition period. The time resolution of the setup was 0.9 ns (FWHM) in the 133 – 482 Kev cascade gate.

The perturbation function for I=5/2 intermediate state is described as [18]:

$$G_2(t) = a_0 + a_1 \cos \omega_1 t + a_2 \cos \omega_2 t + a_3 \cos \omega_3 t$$

The three precession frequencies are:

$$\omega_1 = (E_2 - E_3)\omega_Q$$
$$\omega_2 = (E_1 - E_2)\omega_Q$$
$$\omega_3 = (E_1 - E_3)\omega_Q$$

Where the quadrupole frequency is:

$$\omega_Q = \frac{eQV_{zz}}{40\hbar}$$

$E_1$, $E_2$ and $E_3$ are the three energy levels in which the intermediate state I=5/2 of the probe is split into by the interaction with the EFG and $Q$ is the quadrupole moment of the intermediate state of the nucleus.

Probes occupying more than one site (say, '$i$' sites) gives rise to a total perturbation function with different amplitudes $a_i$

$$G_2(t) = \sum_{i=1}^{n} a_i G_2^i(t)$$

Experimental data were fitted with this function modified by the finite distribution of $\omega_Q$. In the present case we have used lorentzian distribution. Prompt resolution with a Gaussian distribution has also been incorporated in the final form of the perturbation function.

**Experimental Results**

AFM picture of the ball-milled $TiO_2$ particles is shown in Fig-1. Fig-1a shows the size distribution which indicates a mean size of 70 nm. XRD for this sample and those annealed at different temperatures are shown in Fig-2. The widths of the anatase peak at different stages of annealing are shown in Table-1. Quadrupole frequencies ($\omega_Q$) and the asymmetry parameters ($\eta$) for bulk $TiO_2$ doped with [181]Hf in anatase and rutile modifications are 44.01(3) Mrad/s & 0.22(1) and 130.07(9) & 0.56(1) respectively [15]. These TDPAC parameters are required to identify the two phases evolved at different stages of the 70 nm $TiO_2$ samples annealed at different temperatures.

TDPAC spectra for $TiO_2$ sample annealed at different temperatures and for different duration are shown in Fig-3 with the individual spectrum denoted as S1, S2 and so on. TDPAC spectrum for the 70 nm $TiO_2$ adsorbed with [181]Hf and dried at 373K is shown in S1. Corresponding frequency spectrum on the right shows a broad distribution of the frequency. Same sample was annealed at 823K for 4 hrs and the corresponding TDPAC spectrum is shown in S2. The above experiment on the initially anatase phase was repeated with enhancing the time of annealing. S3 shows the TDPAC spectrum for the sample annealed at 823K for 8 hrs. It is known [17] that anatase is converted into rutile at 1123K. A part of the anatase TiO$_2$ was converted into rutile and the tracer $^{181}$Hf was adsorbed on this rutile sample and it was annealed at 1123K for 4 hrs. TDPAC spectrum of this sample is shown in S4. To compare the diffusion of $^{181}$Hf through the rutile surface in sample S4 and the movement of the tracer during the phase transition, the sample S2 was annealed at 1123K for 4 hrs. TDPAC results for this sample are shown in S5. Initially anatase TiO$_2$ adsorbed with $^{181}$Hf tracer was annealed at 1123K for 8 hrs to get a qualitative idea about the time of the conversion process. The results for this sample are shown in S6. Another sample when annealed at 1223K for 8 hrs was found to be completely converted into rutile. This result is shown in S7.

**Discussion**

In this paper we address some aspects of TiO$_2$ crystallite with particular reference to the phase transition from anatase to rutile. First, we measured the hyperfine parameters viz. $\omega_Q$, $\eta$, for the bulk anatase and rutile phases to identify these phases formed in the TiO$_2$ on which the nuclear probe had been adsorbed and annealed at different temperatures. In our earlier experiment [17] we found that TiO$_2$ prepared by the hydrolysis of Ti-isopropoxide and annealed at 823K remains in the anatase phase and the transformation from anatase to rutile takes place only at 1123K. In this present work same conditions were maintained. Here the nuclear probe used is $^{181}$Ta. The results for the anatase phase were measured by us for the first time [15]. On the other hand, that for the rutile is a repetition. This was measured by others. However the method of sample preparation in our case is different from that in ref [16].

XRD and AFM measurements indicate that the starting TiO$_2$ is anatase with an average size of 70nm. We found significant adsorption (~90%) of $^{181}$Hf tracer on this nano TiO$_2$ surface from the neutral solution of the tracer. Hafnium tracer in neutral medium is expected to exist as positively charged poly ionic form similar to that of zirconium [19], which is adsorbed on the surface of TiO$_2$. It may thus be expected that the surface of TiO$_2$ is negatively charged. Detailed study of this adsorption process through the measurement of the enthalpy, adsorption isotherm is in progress.

TDPAC study of this adsorbed sample, S1 (dried at 373K to remove moisture) showed a broad distribution of the quardrapole frequency. This is a clear indication of static inhomogeneity of the $TiO_2$ surface. A possible explanation for this inhomogeneity is expected to have an origin in the uneven arrangement of oxygen bridging sites [20] in $TiO_2$ nano particles. The same sample when annealed at 823K for 4 hours does not show much difference in the TDPAC spectrum (S2) except the reduction in the width of frequency distribution. This indicates the partial removal of the surface inhomogeneity by heating process. Anther explanation for this enhanced homogeneity is the increased surface smoothness due to growth of particle size at higher temperature. TDPAC results indicate that $^{181}Hf$ tracer remains on the surface only till 4 hrs of heating at 823K. XRD and TDPAC results indicate that there is neither any phase transition nor any diffusion of the tracer in the matrix after annealing the sample at 823K for 4 hrs. Width measurement of the anatase peak in the XRD spectrum, shown in table-1 indicates that there is a steady decrease in the width. This clearly indicates the growth of the crystallite. However, no phase transition did take place till four hours of annealing at 823K.

A different behavior, however, was found when this sample was annealed at 823K for 8 hours (S3). The broadening ($\delta$) is removed completely and the $TiO_2$ crystallite is transformed into rutile with a small contribution of anatase. But a large fraction for the probe could not be accounted for. Only a small fraction (25%) of the total anisotropy was identified as rutile plus anatase out of which 95% is rutile and only 5% is anatase. There is a fast decrease in anisotropy. This decrease is yet to be understood. This observation leads to few conclusions: first, the kinetics of phase transition or the mass transfer from surface to bulk is slow. The question, however, arises why almost no probe is found in anatase part. In this experiment with bulk $TiO_2$, we did not see any phase transformation at 823K. It is not unusual for nano particle to undergo phase transformation at lower temperature. During this process of phase change the probe enters into the bulk from the surface and goes preferably to rutile phase. The other possibility is that the rutile phase might have been formed first and then the probe diffuses into the different phases. Both the processes mentioned above might be operating simultaneously. This conclusion is made on the basis of following results.

Initially anatase $TiO_2$ (70 nm), before adsorption of the probe on its surface was annealed at 1123K to convert it into rutile. This temperature was selected on the basis of our work [17] on bulk sample. Then $^{181}$Hf was adsorbed on this supposedly rutile surface. There was much less adsorption of the probe than on the starting anatase $TiO_2$ due to lowering in surface area resulting from annealing. This probe adsorbed $TiO_2$ was annealed at 1123K for 4 hrs again to check the surface to bulk mass transfer. A significant amount of probe was found to be in rutile phase and nothing in anatase phase (S4). It means either there is no anatase phase existing in the sample when annealed at 1123K or $^{181}$Hf does not diffuse into anatase phase. However the subsequent experimental results indicate that the first comment on the nonexistence of anatase is not true. $TiO_2$ (anatase) adsorbed with $^{181}$Hf when annealed at 1123K for 4 hrs, though majority fraction of probe goes into rutile (62%) a significant amount (24%) goes into anatase as well. This indicates the existence of anatase when $TiO_2$ was annealed at 1123K (S5). But the probe was not found in anatase in S4. In S5, thus the existence of probe in anatase phase is not due to diffusion. Had this been so, a significant amount of the probe could be observed in anatase phase in S4 as has been found in S5. The phase transition in 70nm $TiO_2$ thus can be modeled in the following way. The surfaces of nano particles roll over each other during annealing causing growth in the crystallite and the $^{181}$Hf/$^{181}$Ta probe from the surface gets transferred to the bulk which transforms into rutile and partly remains as anatase. Probe in rutile can be either during the process of phase transition or due to diffusion after the rutile is formed or both.

**Conclusion**

Phase transition of 70nm $TiO_2$ takes place at lower temperature than that for the bulk. Annealing at 823K for 4 hrs causes the growth in the crystallite but there was no phase transition. Annealing at the same temperature for longer time leads to phase transition indicating that the kinetics of the phase transition is not fast. However, at higher temperature, phase transition is faster. During phase transition, there occurs growth of the crystallites along with the transfer of the tracer from surface to bulk. This diffusion was found to take place in rutile, but not in anatase. Thus the surface to bulk mass transfer plays a significant role in the phase transition.

**Acknowledgement**

The authors are thankful to Mr. R.K. Chatterjee, Radiochemistry Laboratory, VECC, for his help in preparing the samples and data taking. The authors are indebted to Dr. P.K. Pujari, Head, Nuclear Chemistry Section, Radiochemistry Division, Dr. V.K. Manchanda, Head, Radiochemistry Division and Dr. S.K. Aggarwal, Head, Fuel Chemistry Division and Dr. Meera Venkatesh, Head Radiopharmaceuticals Division, Bhabha Atomic Research Centre, for their keen interest and support in this work.

Table-1: Width of the XRD peak for the anatase $TiO_2$ samples annealed at different temperatures

| Sample | Centre(2θ degrees) | Width(2θ degrees) |
|---|---|---|
| Bulk $TiO_2$ | 25.287 | 0.172 |
| 70 nm $TiO_2$ | 25.248 | 0.357 |
| 70 nm $TiO_2$ annealed at 823K for 4 hrs | 25.303 | 0.262 |
| 70 nm $TiO_2$ annealed at 1123K for 4 hrs | 25.301 | 0.150 |

Table-2: TDPAC parameters for different samples

| Sample | $\omega_Q$ (Mrad/sec) | | η | | δ (%) | | Composition (%) | |
|---|---|---|---|---|---|---|---|---|
| | Anatase | Rutile | Anatase | Rutile | Anatase | Rutile | Anatase | Rutile |
| S1 | 82.9(4) | - | 0.36(11) | - | 32.0(2) | - | 100 | 0 |
| S2 | 77.5(5) | - | 0.62(7) | - | 25.2(5) | - | 100 | 0 |
| S3 | 44.6(2) | 132.4(1) | 0.30(2) | 0.53(2) | 0.0(1) | 7.9(1) | 5 | 95 |
| S4 | - | 134.0(7) | - | 0.51(1) | - | 5.2(5) | 0 | 100 |
| S5 | 44.0(6) | 130.9(3) | 0.27(4) | 0.54(2) | 6.2(3) | 1.1(6) | 21 | 79 |
| S6 | 44.0(7) | 130.3(1) | 0.30(5) | 0.55(1) | 0.1(1) | 0.3(1) | 8 | 92 |
| S7 | - | 130.5(1) | - | 0.55(1) | - | 1.0(1) | 0 | 100 |

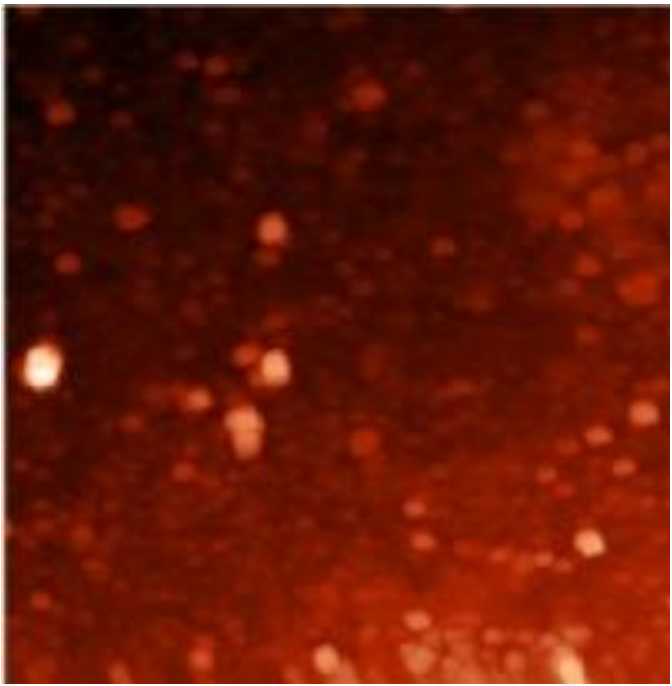

**Fig-1:** AFM picture of ball-milled TiO$_2$ in anatase form

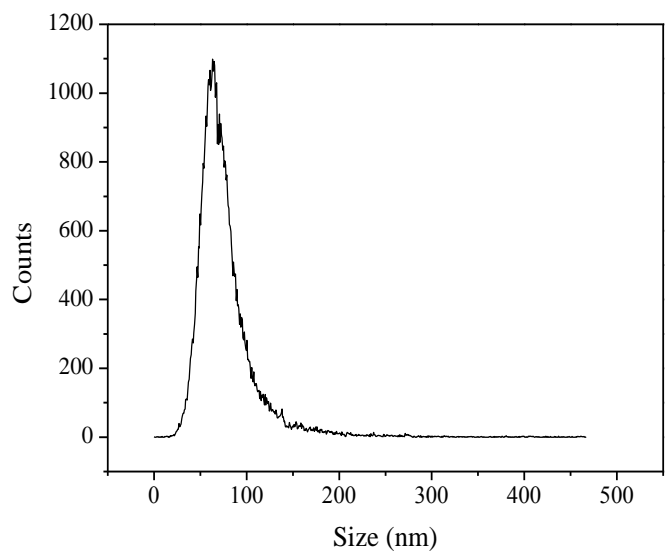

**Fig-1a:** Size distribution of the TiO$_2$ particles shown in fig-1

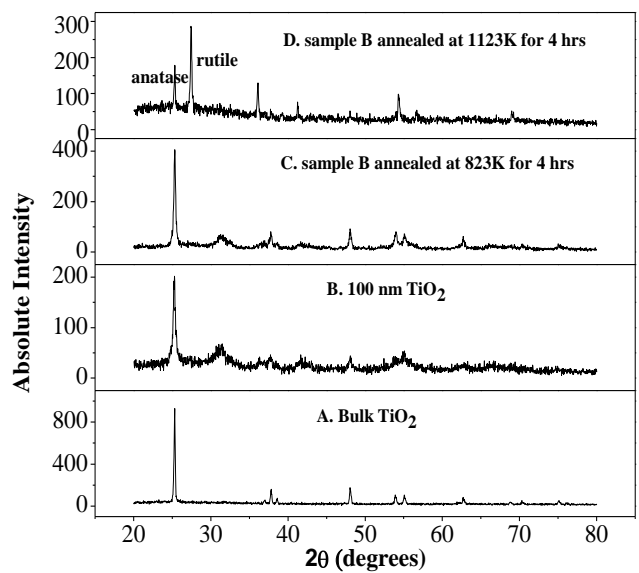

**Fig-2:** XRD spectra of TiO$_2$ annealed at different temperatures

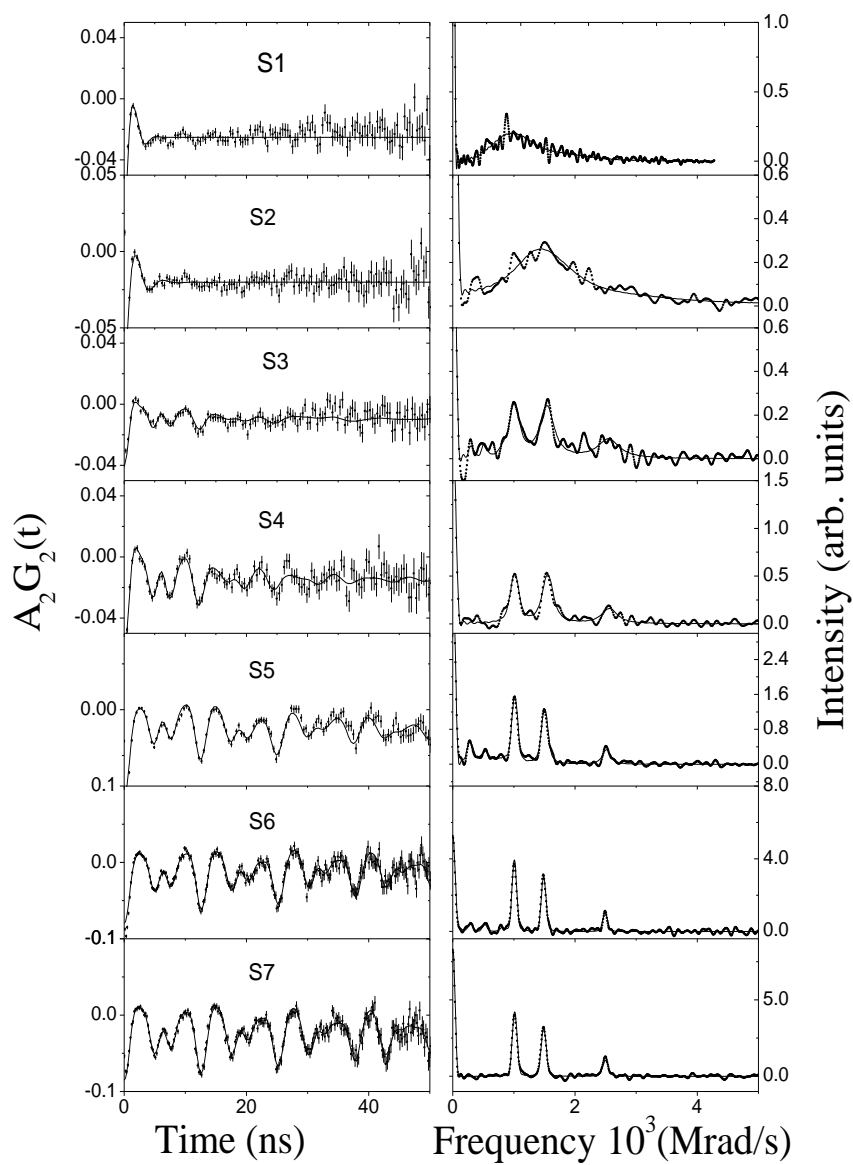

**Fig-3**: TDPAC spectra of nano $TiO_2$ samples annealed at different temperatures; **S1**: $^{181}$Hf tracer adsorbed on 70nm $TiO_2$ (anatase) dried at 373K. **S2**: S1 annealed at 823K for 4 hrs. **S3**: S1 annealed at 823K for 8 hrs. **S4**: $^{181}$Hf tracer adsorbed on TiO2 (rutile) annealed at 1123K for 4 hrs. **S5**: S2 annealed at 1123K for 4 hrs. **S6**: S1 annealed at 1123K for 8 hrs. **S7**: S1 annealed at 1223K for 8 hrs.